\documentclass[twocolumn,aps,prb]{revtex4}
\usepackage[T1]{fontenc}
\usepackage[latin1]{inputenc}
\usepackage{amsmath}
\usepackage{amstext}
\usepackage{epsfig}
\usepackage{epic}
\usepackage{eepic}
\usepackage{amssymb}
\usepackage{exscale}
\usepackage{pst-node,array}
\begin{document}
\title{Phase transitions in the Shastry-Sutherland lattice}
\author{Mohamad Al Hajj and Jean-Paul Malrieu}
\affiliation{Laboratoire de Physique Quantique, IRSAMC/UMR5626, Universit\'e Paul Sabatier, 118 route 
de Narbonne, F-31062 Toulouse Cedex 4, France}
\begin{abstract}
Two recently developed theoretical approaches are applied to the Shastry-Sutherland lattice, varying the 
ratio $J'/J$ between the couplings on the square lattice and on the oblique bonds. A self-consistent 
perturbation, starting from either Ising or plaquette bond singlets, supports the existence of an 
intermediate phase between the dimer phase and the Ising phase. This existence is confirmed by the results 
of a renormalized excitonic method. This method, which satisfactorily reproduces the singlet triplet 
gap in the dimer phase, confirms the existence of a gapped phase in the interval $0.66<J'/J<0.86$
\bigskip
\end{abstract}
\maketitle
\section{Introduction}
Strongly correlated electron systems, and in particular those which can be treated as quantum magnets, are the subject of 
a continuous and intense attention from experimentalists and theoreticians. They frequently exbibit complex phase diagrams, 
and non-trivial low energy physics. Among the spin lattices those presenting spin frustrations both present unconventional
phases and quantum phase transitions and represent a challenge for theoreticians. The quantum Monte Carlo methods, which 
may be considered as the most reliable treatment for non-frustrated 2-D or 3-D lattices, cannot be applied in these cases.
Among the 2-D frustrated lattices one may quote the checkerboard, the Kagome and the triangular lattices, which have been
the subject of numerous theoretical works.   
The $SrCu_2(BO_3)_2$ lattice \cite{Ref1} is a famous two-dimensional anti-ferromagnetic system presenting a spin gap and free from long range order. The Copper atoms are of $d^9$ character and can be seen as $S=1/2$ spins. 
This lattice may be considered as a realization of the Shastry-Sutherland model \cite{Ref2} which can be schematized 
as a square lattice, with $J'$ anti-ferromagnetic coupling between nearest neighbors, and diagonal anti-ferromagnetic 
interactions $J$ in one plaquette over two, as pictured in Fig. \ref{fig1}. This system is supposed to obey the 
corresponding Heisenberg Hamiltonian
\begin{equation}
H=J'\sum_{[i,j]}\overrightarrow{S_i}\overrightarrow{S_j}+J\sum_{\langle i,j \rangle}\overrightarrow{S_i}\overrightarrow{S_j}
\end{equation}
where the couples $[i,j]$ concern the bonds of the square lattice and ${\langle i,j \rangle}$ the connected 
pairs of next nearest neighbor atoms. The real material has been the subject of intense experimental studies,
showing the existence of a spin gap,\cite{Ref3} an almost localized nature of the singlet triplet 
excitation,\cite{Ref4} and the existence of magnetization plateaux.\cite{Ref5,Ref6} The Shastry-Sutherland Hamiltonian has been widely studied by theoreticians
(for review see S. Miyahara and K. Ueda \cite{Ref7}), varying the $J'/J$ ratio. For small values of $J'/J$ the 
ground state 
is a product of singlets strictly localized on the oblique bonds. The real material would correspond to a ratio
$J'/J=0.635$, close to the critical value where this phase disappears. Oppositely when $J$ is small, the 
perturbed 2-D square lattice is gapless with long range order. Early studies based on exact 
diagonalizations,\cite{Ref8} Ising expansion\cite{Ref9} or dimer expansions,\cite{Ref10} and real space 
renormalization group with effective interaction (RSRG-EI) \cite{Ref11} predict a simple transition between the Ising
phase and the dimer phase for $J'/J=0.696$. Other works have suggested the existence of an 
intermediate phase. From Schwinger  boson mean field theory \cite{Ref12} this phase would be an helical 
ordered state, ranging between $J'/J=0.6$ and 0.9 while a plaquette expansion \cite{Ref13,Ref14} predicts a 
plaquette singlet based phase for $0.677<J'/J<0.86$, a result supported by other exact diagonalization 
results.\cite{Ref15} Weihong et al.\cite{Ref16} suggested that the intermediate phase between $J'/J=0.69$ 
and 0.83 might be columnar rather than plaquette singlet.

The present paper employs two recently developed methods to study the phase diagram of this lattice. 
We first will consider the cohesive energy of the lattice, using a self consistent pertubation (SCP) \cite{Ref17} 
which may be seen as a modified coupled cluster expansion.\cite{Ref18,Ref19,Ref20,Ref21} The method starts from a reference 
function which can be the Ising spin distribution or a columnar product of bond singlets. It happens that the cohesive 
energy calculated from the latter reference function is the lower one for $0.66<J'/J<0.86$, the lowest energy being 
obtained from the Ising reference for $J'/J>0.86$. The calculated cohesive energies obtained by the RSRG-EI 
with $(3\times3)$ sites square blocks or by the contractor renormalization (CORE)\cite{Ref22} 
with columnar blocks of $(2\times6)$ sites agree with the values obtained from SCP and support the idea of an intermediate 
phase.

A second section concentrates on the calculation of the gap, using a renormalized excitonic method (REM), based on the 
scale-change ideas inspiring the renormalization group methods. The methods proceeds through the definition of blocks 
with even number of sites.
The design of the relevant blocks is different for the dimer phase and for the other phases. The calculated gap for the 
dimer phase is in good agreement with that of previous theoretical study \cite{Ref13,Ref14} and compatible with 
experimental evidences.\cite{Ref7} Approaching the problem from columnar blocks one obtains a finite gap for 
$J'/J<0.88$. This result confirms the existence of an intermediate phase.
\section{Calculation of the cohesive energy} 
If one defines $\lambda=J'/(J'+J)$ and take $J'+J$ as the energy unit, the product of singlets on the oblique dimers
($J$ interactions) is an eigenfunction whatever $\lambda$, and its cohesive energy is $J'+J=1$. 
\subsection{Self-consistent perturbation}
We have used the recently 
proposed SCP method \cite{Ref17} to evaluate the cohesive energy of other 
phases. The method can be seen as a modified coupled cluster method.\cite{Ref18,Ref19,Ref20,Ref21} It starts from a 
reference function $\Phi_0$, supposed to be a relevant zero-order function for the considered phase. This 
function is highly localized. In practice $\Phi_0$ will be 
\begin{itemize}
\item[-] the Ising spin distribution on the square 2-D lattice,
\item[-] or a product of bond singlets in a columnar arrangement. On each of these bonds one may 
also define a local triplet state.
\end{itemize}
The action of $H$ on $\Phi_0$ generates a first generation of local excited function $\Phi_i$
($\langle\Phi_i\vert H\vert\Phi_0\rangle\not=0$). In the so-called intermediate normalization convention
\begin{equation}
\Psi=\Phi_0+\sum_iC_i\Phi_i+\sum_{\alpha}C_{\alpha}\Phi_{\alpha},
\end{equation}
where the vectors $\Phi_{\alpha}$ do not interact with $\Phi_0$ ($\langle \Phi_{\alpha}\vert H\vert\Phi_0\rangle=0$),
the knowledge of the coefficients of the first generation $\Phi_i$'s is sufficient to fix the ground state 
energy
\begin{figure}[t]
\centerline{\includegraphics[scale=0.6]{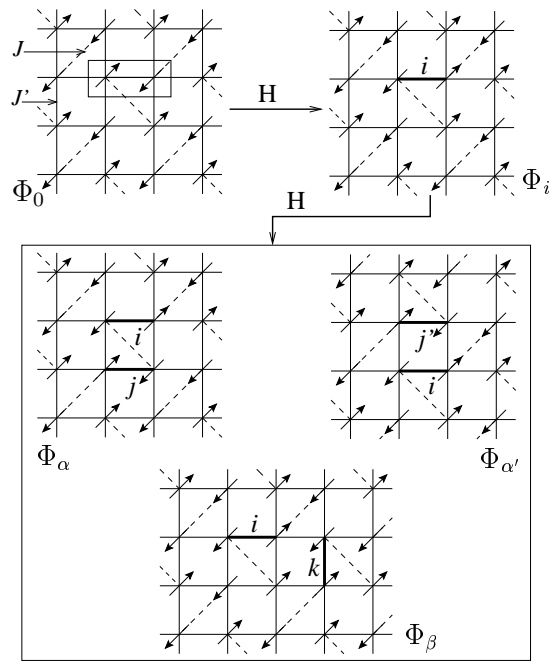}}
\caption{Genealogy of the wave function starting from Ising $\Phi_0$. $\Phi_0$ = reference function,
$\Phi_i$ = first generation function, $\Phi_{\alpha}$, $\Phi_{\alpha'}$, $\Phi_{\beta}$ types of non
factorisable second generation functions.}
\label{fig1}
\end{figure}
\begin{equation}
E=\langle\Phi_0\vert H\vert\Phi_0\rangle+\sum_iC_i\langle\Phi_0\vert H\vert\Phi_i\rangle.
\end{equation}
Starting from Ising $\Phi_0$, the vectors $\Phi_i$ are obtained by a spin exchange on any bond of the 
2-D square lattice (Fig. \ref{fig1}). Starting from the product $\Phi'_0$ of singlets on the $X$-directed bonds in a 
columnar arrangement (Fig. \ref{fig2}), the vectors $\Phi_i$ are excited 
singlet states produced as products of two triplets on interacting bonds. They are of four different types
in a Shastry-Sutherland lattice, as pictured in Fig. \ref{fig2}. 

The determination of the coefficients $C_i$ is governed by the corresponding eigenequation, adopting the 
compact notation $H_{ij}=\langle \Phi_i \vert H \vert \Phi_j \rangle$,
\begin{equation}
(H_{ii}-E)C_i+H_{i0}+\sum_j H_{ij}C_j+ \sum_{\alpha\not \in S_1} H_{i\alpha}C_{\alpha}=0.
\end{equation}
An estimation of the coefficients of the second generation functions $\Phi_{\alpha}$ (such that $H_{i\alpha}$)is only 
necessary for those which are obtained from $\Phi_i$ by operations in the strict neighborhood of the bonds involved in the 
process $T_i^+$ creating $\Phi_i$ from $\Phi_0$ ($\Phi_i=T_i^+\Phi_0$). The spin exchanges or excitations $T_k^+$
on remote bonds cancel, according to the linked cluster theorem, since in that case
\begin{equation}
C_{T_k^+ \vert \Phi_i\rangle}=C_kC_i.
\end{equation}
The number of second generation functions to be considered is therefore very limited. Their coefficient is 
estimated from the $C_i$'s according to perturbative arguments, with a systematic consideration of exclusion 
principle violating corrections of the energy denominators, which introduce infinite summations of diagrams 
and speed the convergence. Details of the method are given elsewhere.\cite{Ref17} The equations for the precise problem
are given in the Appendix, the results appear in Fig. \ref{fig3}. One sees that
\begin{itemize}
\item[-] the energy calculated from Ising $\Phi_0$ is the lowest one for large values of $J'/J$. This energy curve cuts 
the energy of the localized dimer phase for $J'/J=0.69$. This value compares quite
well with other estimates,\cite{Ref8,Ref9,Ref10,Ref11} and in particular with that of a RSRG-EI\cite{Ref11} with 
$(3\times3)$ sites square blocks ($(J'/J)_c=0.696$)
\item[-] however the energy calculated from the columnar arrangement of bond singlets on the 2-D square 
lattice happens to be lower than the preceding one for $J'/J<0.859$. The corresponding energy is the lowest 
one for $0.661<J'/J<0.859$. Hence these calculations support the suggestion of the existence of three phases,
with an intermediate phase between the dimer phase and the Ising phase.
\end{itemize}
\begin{figure}[t]
\centerline{\includegraphics[scale=0.6]{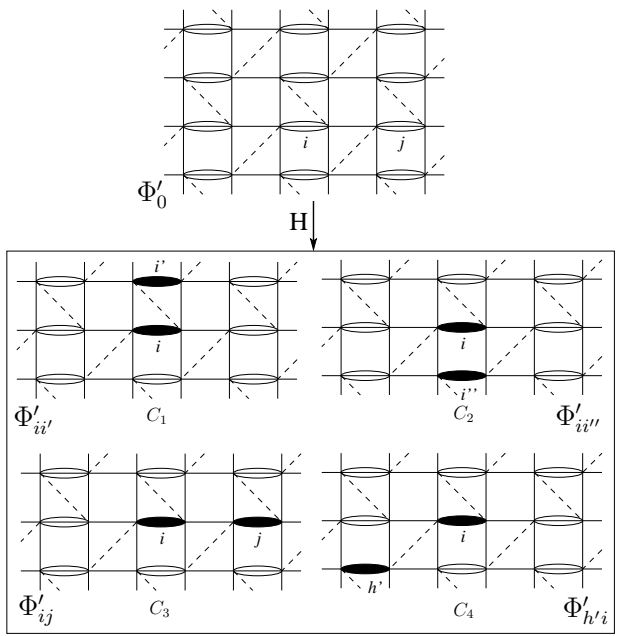}}
\caption{Genealogy of the wave function starting from columnar product of bond singlets empty ellipses. 
Dark ellipses represent local triplets. The figure pictures the 4 types of first generation singlet states. The non
factorisable second generation functions are too numerous to be pictured.}
\label{fig2}
\end{figure}
\begin{figure}[b]
\centerline{\includegraphics[scale=0.38]{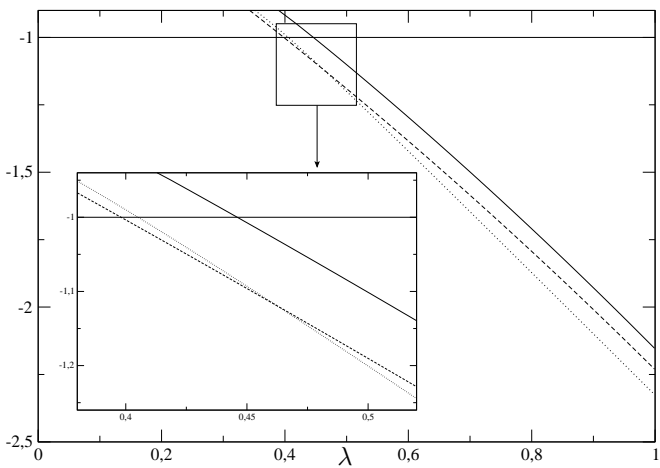}}
\caption{Cohesive energy (with the unit $(J+J')$) as a function of the $\lambda$ parameter. The horizontal straight line 
concerns the dimer phase, - - - SCP expansion from columnar bond singlets $\Phi'_0$, $\cdots$ SCP expansion from Ising 
$\Phi_0$, --- SCP expansion from shifted bond singlets $\Phi''_0$.}
\label{fig3}
\end{figure}
However the fact that one finds two distinct values of the energy from two distinct reference functions within
an approximate algorithm does not prove the existence of two phases. It is possible that at convergence the 
two wave operators, $\Omega$ from $\Phi_0$ and $\Omega'$ from $\Phi'_0$, lead to the same wave function 
\begin{equation}
\Psi_0=\Omega\Phi_0=\Psi'_0=\Omega'\Phi'_0.
\end{equation}
As an argument in favor of two distinct phases we may mention that starting from an other function 
$\Phi''_0$, energetically degenerate with $\Phi'_0$  
($\langle \Phi''_0 \vert H \vert \Phi''_0 \rangle = \langle \Phi'_0 \vert H \vert \Phi'_0 \rangle$), 
product of bond singlets on parallel bonds which do not belong to the same plaquette (see Fig. \ref{fig4}),
the calculated energy always remains above that obtained from Ising $\Phi_0$,  as seen in Fig. \ref{fig3}. 
It is not easy, at least from our SCP approach, to determine wether the intermediate phase is based on a columnar 
arrangement of bond singlets or is a plaquette phase. The second one keeps isotropic properties in the $X$ and $Y$ 
directions, while the former one breaks this symmetry. It is interesting to remark that our reference wave function 
$\Phi'_0$ is strongly anisotropic, with a probability zero to find the parallel spins in the $X$-directed bonds of a 
plaquette and a probability 1/2 to find parallel spins in the $Y$-directed bonds. After consideration of the possible 
excitations, the probability to find parallel spins along the two $X$-directed bond of an empty plaquette becomes 
$87\%$ of the probability to find parallel spins on the $Y$-directed bonds of the same plaquette. The isotropy is 
almost restaured by the self-consistent evaluation of the excitation amplitudes. The result, obtained from a highly broken-symmetry reference function, pleads in favor of a plaquette phase.
\begin{figure}[h]
\centerline{\includegraphics[scale=0.6]{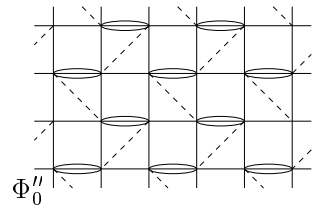}}
\caption{Alternative reference function $\Phi''_0$.}
\label{fig4}
\end{figure}
\subsection{Renormalization group methods}
Renormalization group techniques have also been employed. 
A preceding work \cite{Ref11} had considered square 
9 sites i.e., $(3\times3)$ sites blocks, the ground state of which is an $S_z=\pm1/2$ doublet, i.e., a quasi spin.
The lattice of blocks is isomorphic to the Shastry-Sutherland lattice, with two types of effective interactions $J'^{(1)}$ 
and $J^{(1)}$ between the blocks, the amplitude of which can be obtained from the exact spectrum of the 
dimers of blocks, according to the effective Hamiltonian theory.\cite{Ref23} The ratio $J'^{(1)}/J^{(1)}$
is larger than $J'/J$ for $J'/J>0.696$. For $J'/J=0.696$, $J'^{(1)}/J^{(1)}=J'/J$, i.e., this value is a fixed point, for 
which the cohesive energy calculated by iterating the RSRG-EI procedure practically coincide with that of the dimer 
phase. The behaviour of the so-calculated cohesive energy of the Ising phase is plotted in Fig. \ref{fig6}.
The intersection of the cohesive energy curves of the dimer phase and of the RSRG-EI estimate of the Ising phase occurs 
for $J'/J=0.672$.\cite{Ref11} 
The method does not produce an intermediate phase, when it is applied to these isotropic square blocks.

If one uses 2n sites blocks, with a non degenerate singlet ground state $\psi_A^0$ 
of energy $E_A^0$, the ground state of the lattice is studied from the product of the ground state of the blocks
$\Psi^0=\prod_A\psi_A^0$ according to the simplest version of the CORE method.\cite{Ref22} 
The knowledge of the ground state exact energy $E_{AB}^0$ of the $AB$ dimers enables one to define an effective interaction 
$v_{AB}$ between the blocks\cite{Ref22} as 
\begin{table}[t]
\caption{Cohesive energy (with the unit $(J+J')$) for intermediate values of the $J'/J$ ratio.}
\begin{ruledtabular}
\begin{tabular}{lcccc}
$J'/J$   & SCP(Ising) & SCP(Plaquette)  & $\mbox{CORE}^{\mbox{a}}$ & $\mbox{RSRG-EI}^{\mbox{b}}$   \\
\hline
0.65016  & -0.97804    & -0.99257       & -0.99803 & -0.98501 \\ 
0.66667  & -0.98987    & -1.00355       & -1.00787 & -0.99609 \\
0.69461  & -1.00989    & -1.02191       & -1.02488 & -1.01490 \\
0.72413  & -1.03026    & -1.04034       & -1.04251 & -1.03362 \\
0.81818  & -1.09291    & -1.09604       & -1.09783 & -1.09510 \\
0.85185  & -1.11419    & -1.11475       & -1.11720 & -1.11790 \\
1.0      & -1.20061    & -1.19028       & -1.19478 & -1.20920 \\
\hline
\multicolumn{5}{l}{(a) from $(2\times6)$ sites columnar blocks}     \\
\multicolumn{5}{l}{(b) from $(3\times3)$ sites square blocks (cf. ref. 11).}   \\
\end{tabular}
\end{ruledtabular}
\end{table}
\begin{figure}[b]
\centerline{\includegraphics[scale=0.6]{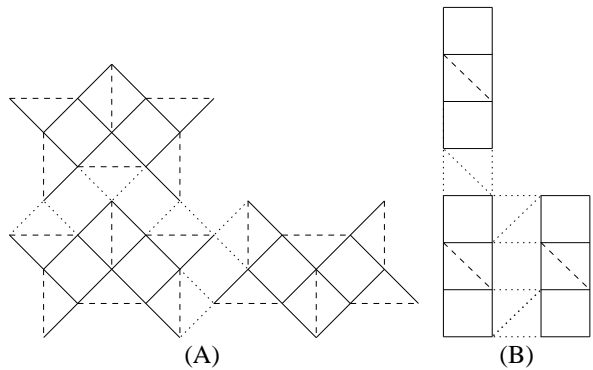}}
\caption{(A) definition of the periodizable 12 sites blocks for the calculation of the excitation gap in the
dimer phase, (B) definition of $(2\times2n)$ sites columnar blocks for the calculation of the excitation gap in the
plaquette phase, and application of the CORE method.}
\label{fig5}
\end{figure}
\begin{equation}
v_{AB}=E_{AB}^0-E_{A}^0-E_{B}^0
\end{equation}
and the energy per block is 
\begin{equation}
E_{A}=E_{A}^0+\frac{1}{2}\sum_Bv_{AB}.
\end{equation}
In the research of a tentative columnar phase, we have considered 12 $(2\times 6)$ sites columnar blocks 
built from three aligned plaquettes. Interestingly enough the calculated cohesive energy is the lowest one 
(cf Fig. \ref{fig6}) in the interval $0.656<J'/J< 0.845$ (RSRG-EI) or 0.901 (SCP). This result is in excellent agreement 
with our SCP result, and supports the existence of an intermediate plaquette phase. Table 1 reports the calculated 
values of the cohesive energy in the problematic domain of the $J'/J$ ratio one sees the good agreement between the SCP 
from Ising and RSRG-EI on one hand and between the SCP from columnar arrangement of bond singlets and CORE with 
$(2\times6)$ sites columnar blocks on the other hand. 
\begin{figure}[t]
\centerline{\includegraphics[scale=0.98]{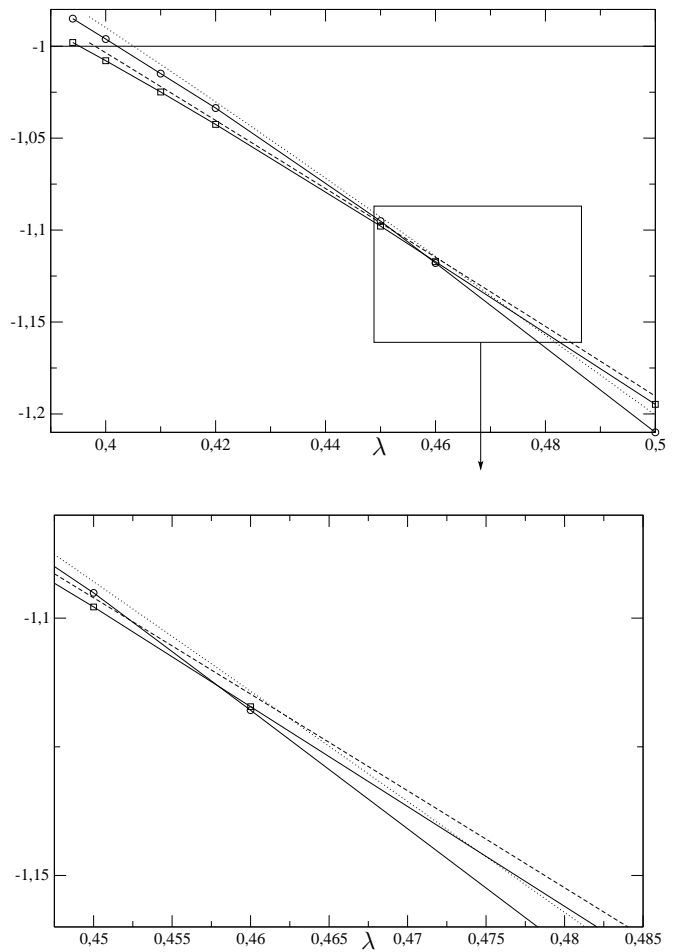}}
\caption{Evolution of the cohesive energy (with the unit $(J+J')$) as a fonction of $\lambda=J'/(J+J')$. The horizontal 
line concerns the dimer phase,  - - - SCP from columnar bond singlets $\Phi'_0$, $\cdots$ SCP from Ising $\Phi_0$, 
---\hspace{-1.0mm}$\circ$\hspace{-1.0mm}--- RSRG-EI from $(3\times3)$ sites square blocks,\cite{Ref11} 
---\hspace{-1.3mm}$\Box$\hspace{-1.3mm}--- CORE with $(2\times6)$ sites columnar 
blocks. The lower part is a zoom in the region of crossing between the intermediate and Ising phases.}
\label{fig6}
\end{figure}
\section{Calculation of the singlet triplet gap}
The study of the singlet triplet gap may bring additional informations. It has been performed 
according to an other renormalization group technique, the REM,\cite{Ref24} which starts from blocks $A$
with non degenerate ground state singlet $\psi^0_A$ and an excited triplet state $\psi_A^{\ast}$, and 
build the excited state from linear combinations of locally excited states, $\Psi_A^{\ast}=
\psi_A^{\ast}\prod_B\psi_B^0$. The knowledge of the exact spectrum of the dimers of blocks makes possible the 
calculation of the effective interaction $v_{(A^{\ast})B}$ between an excited triplet and neighbor singlets
and of an effective excitation-hopping integral $h_{AB}$ which couples $\psi_A^{\ast}\psi_B^0$ with 
$\psi^0_A\psi_B^{\ast}$.

\begin{figure}[b]
\centerline{\includegraphics[scale=0.72]{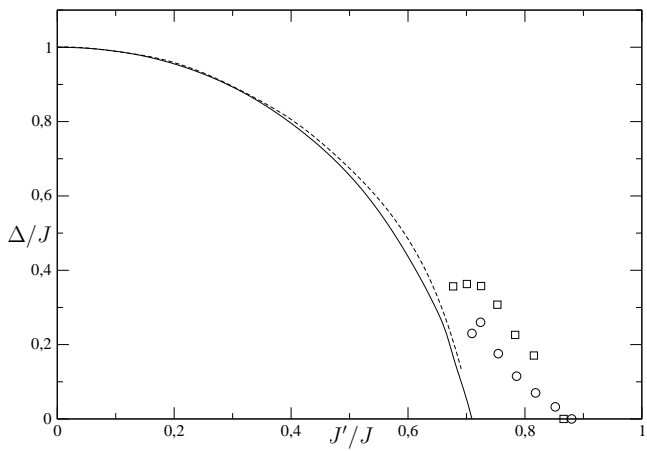}}
\caption{Dependence of the singlet-triplet gap on the $J'/J$ ratio. --- REM from 12-site blocks
(Fig. \ref{fig5}(A)) for the dimer phase, - - - results of a dimer expansion,\cite{Ref9} $\circ$ extrapolated REM 
from columnar plaquette blocks (Fig. \ref{fig5}(B)), $\Box$ results of a plaquette expansion.\cite{Ref13,Ref14}}
\label{fig7}
\end{figure}

We first have applied the method to the dimer phase. It is known that in this phase the triplet 
excitation on a given dimer bond can only propagate to a neighbor one through a 6th-order 
process involving a vortex of 4 dimers. A complete evaluation of the interaction leads to 
\centerline{\includegraphics[scale=1.0]{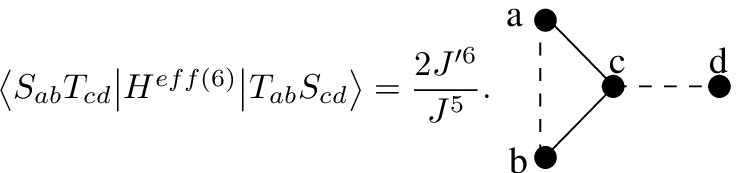}} \\
This interaction remains small when $J'$ increases, and the energy gap slowly deviates from $J$. A 
convenient design of 12 sites (6 dimers bonds) blocks involving two vortex have been pictured in Fig. \ref{fig5}(A).
The gap calculated from the REM are reported in the left part of Fig. \ref{fig7}, together with those of a 
dimer expansion.\cite{Ref9} The excitation energy is tangent to the line $\Delta/J=1$ at the origin, 
as expected, but falls down rapidly near the phase transition. It vanishes for $J'/J=0.7006$. For the 
$J'/J=0.635$, i.e., the value proposed for the real material, our calculation gives 
$\Delta =0.32J$, which satisfactorily compares with the experimental value (30-35 K)\cite{Ref7} 
if one accepts the usually proposed value of $J$ (85 K). 
For the columnar plaquette phase, blocks built from aligned plaquettes have been considered (Fig. \ref{fig5}(B)),
together with an extrapolation on the block size.\cite{Ref24} The results appear in Fig. \ref{fig7} together with those 
of a plaquette expansion.\cite{Ref13,Ref14} One sees that the system is gapped for $J'/J<0.883$, and 
gapless beyond this value. The gap in this phase goes through a maximum for $J'/J \simeq 0.73$, a result also obtained 
by Koga and Kawakami.\cite{Ref13,Ref14} Our results confirm the existence of an intermediate gapped phase. 
\section{Conclusion}
The present work, employing essentially different methods based on either coupled cluster type 
expansions or renormalization group techniques, present consistent results confirming the existence of an 
intermediate phase in the Shastry-Sutherland lattice, and its plaquette or columnar nature. The results regarding the 
location of the 
phase transitions are gathered in Table 2. While the phase transition between the dimer and the Ising phases would 
occur at $J'/J$ between 0.67 and 0.70, the intermediate phase is the lowest one from 
$J'/J=0.656$ (CORE 12 sites) - 0.661 (SCP) to $J'/J=0.859$ (SCP)  - 0.901 (CORE/SCP). Obtained from 
new methods, these results agree with those of refs. 13-16 and reduce the remaining uncertainties concerning the 
existence, domain and nature of the intermediate phase in the Shastry-Sutherland lattice.
\begin{table}[h]
\begin{center}
\caption{Critical values of the $J/J'$ ratio, based on cohesive energy (Coh) or gap calculations (Gap).}
\begin{ruledtabular}
\begin{tabular}{lllc}
               &Criterion     & Methods                      & $J'/J$                        \\                             
\hline
               &Coh           & $\mbox{dimer}^{\mbox{a}}/\mbox{SCP}^{\mbox{b}}$      & 0.690    \\       
Dimer/Ising    &Coh           & $\mbox{dimer}^{\mbox{a}}/\mbox{RSRG-EI}^{\mbox{c}}$  & 0.672    \\
               &Gap           & $\mbox{REM}^{\mbox{d}}$                              & 0.701    \\
\hline
               &Coh          & $\mbox{dimer}^{\mbox{a}}/\mbox{SCP}^{\mbox{e}}$     & 0.661      \\
\raisebox{1.5ex}[0cm][0cm]{Dimer/Interm.}& Coh &$\mbox{dimer}^{\mbox{a}}/\mbox{CORE}^{\mbox{f}}$ & 0.656 \\
\hline            
               &Coh& $\mbox{SCP}^{\mbox{e}}/\mbox{SCP}^{\mbox{b}}$      & 0.859  \\
               &Coh& $\mbox{CORE}^{\mbox{f}}/\mbox{SCP}^{\mbox{b}}$     & 0.901  \\
Interm./Ising  &Coh& $\mbox{SCP}^{\mbox{e}}/\mbox{RSRG-EI}^{\mbox{c}}$  & 0.826  \\
               &Coh& $\mbox{CORE}^{\mbox{f}}/\mbox{RSRG-EI}^{\mbox{c}}$ & 0.845  \\  
               &Gap& $\mbox{REM}^{\mbox{g}}$                            & 0.883  \\      
\hline
\multicolumn{4}{l}{(a) exact energy} \\
\multicolumn{4}{l}{(b) from Ising $\Phi_0$} \\
\multicolumn{4}{l}{(c) from $(3\times3)$ sites square blocks (cf. ref. 11)} \\
\multicolumn{4}{l}{(d) gap vanishing of the dimer phase, 12-site blocks}\\
\multicolumn{4}{l}{(Fig. \ref{fig5}(A))}\\
\multicolumn{4}{l}{(e) from columnar bond singlets $\Phi'_0$} \\
\multicolumn{4}{l}{(f) from $(2\times6)$ sites columnar blocks} \\
\multicolumn{4}{l}{(g) gap vanishing of the Intermediate phase,}\\
\multicolumn{4}{l}{extrapolations from columnar blocks.}\\
\end{tabular}
\end{ruledtabular}
\end{center}
\end{table}
\begin{widetext}
\appendix
\section{}
\noindent SCP equations for the Shastry-Sutherland lattice\\
(1) Equation from Ising function. There is a unique type of spin exchange, and a single coefficient $C$.
\begin{eqnarray}
& & \big[6J'-2J-7J'C\big]C+J'+JC+J'\Bigg[2\Bigg(\frac{12J'-4J-14J'C}{8J'-4J-12J'C}\Bigg)-1\Bigg]C^2+
J'\Bigg[2\Bigg(\frac{12J'-4J-14J'C}{8J'-2J-12J'C}\Bigg)-1\Bigg]C^2 \nonumber \\
& & +4J'\Bigg[\frac{12J'-4J-14J'C}{10J'-2J-13J'C}-1\Bigg]C^2+10J'\Bigg[\frac{12J'-4J-14J'C}{10J'-4J-13J'C}-1\Bigg]C^2
+2J'\Bigg[\frac{12J'-4J-14J'C}{12J'-2J-14J'C}-1\Bigg]C^2 \nonumber \\
& & -\frac{J^2C}{8J'-2J-8J'C}=0,
\end{eqnarray}
\begin{equation}
E_{coh}=-2J'(1-C).
\end{equation}
(2) Equations from bond singlets functions. There are four types of double excitations, as pictured in Fig. 2.
Using the notations
\begin{equation*}
\langle\Phi'_0\vert H\vert\Phi'_{ii'}\rangle=h_1=\frac{\sqrt3\big[2J'-J\big]}{2}, 
\ \langle\Phi'_0\vert H\vert\Phi'_{ii''}\rangle=h_2=\sqrt3 J', 
\ \langle\Phi'_0\vert H\vert\Phi'_{ij}\rangle=h_3=\frac{\sqrt3J'}{2}, 
\ \langle\Phi'_0\vert H\vert\Phi'_{ih'}\rangle=h_4=\frac{\sqrt3J}{2}
\end{equation*}
the four coupled polynomial equations are
\begin{eqnarray}
& & \big[2J'-J-\big(h_1C_1+2h_2C_2+4h_3C_3+2h_4C_4\big)\big]C_1+h_1+\frac{2h_1C_3^2}{3}\nonumber \\
& & -\frac{1}{3}\Bigg[\frac{2h_2\big(h_2C_1+h_1C_2\big)+4h_3\big(2h_3C_1+2h_1C_3\big)+2h_4\big(h_4C_1+h_1C_4\big)}
{4J'-\big(2h_1C_1+2h_2C_2+4h_3C_3+2h_4C_4\big)}\Bigg]=0,
\end{eqnarray}
\begin{eqnarray}
& & \big[2J'-\big(2h_1C_1+h_2C_2+4h_3C_3+2h_4C_4\big)\big]C_2+h_2-\frac{2\big[h_4C_3+h_3C_4\big]}{\sqrt3} 
+\frac{2h_2C_3^2}{3} \nonumber \\
& & -\frac{1}{3}\Bigg[\frac{2h_1\big(h_1C_2+h_2C_1\big)+2h_3\big(2h_3C_2+2h_2C_3\big)}
{4J'-\big(2h_1C_1+2h_2C_2+4h_3C_3+2h_4C_4\big)}\Bigg]=0,
\end{eqnarray}
\begin{eqnarray}
& & \big[3J'-\big(2h_1C_1+2h_2C_2+3h_3C_3+2h_4C_4\big)\big]C_3+h_3-\frac{h_2C_4+h_4C_2}{\sqrt3} 
+\frac{h_3\big[C_1^2+C_2^2+C_4^2\big]}{3}
\nonumber \\
& & -\frac{1}{3}\Bigg[\frac{2h_1\big(2h_1C_3+2h_3C_1\big)+h_2\big(2h_2C_3+2h_3C_2\big)+4h_3^2C_3+h_4\big(2h_4C_3+2h_3C_4\big)}{4J'-\big(2h_1C_1+2h_2C_2+4h_3C_3+2h_4C_4\big)}\Bigg]=0,
\end{eqnarray}
\begin{eqnarray}
& & \big[4J'-J-\big(2h_1C_1+2h_2C_2+4h_3C_3+h_4C_4\big)\big]C_4+h_4-\frac{2\big[h_2C_3+h_3C_2\big]}{\sqrt3} 
+\frac{2h_4C_3^2}{3} \nonumber \\
& & -\frac{1}{3}\Bigg[\frac{2h_1\big(h_1C_4+h_4C_1\big)+2h_3\big(2h_3C_4+2h_4C_3\big)}
{4J'-\big(2h_1C_1+2h_2C_2+4h_3C_3+2h_4C_4\big)}\Bigg]=0,
\end{eqnarray}
\begin{equation}
E_{coh}=\frac{\big[-(7J'+J)+h_1C_1+h_2C_2+2h_3C_3+h_4C_4\big]}{4}.
\end{equation}
\end{widetext}

\end{document}